\documentclass[journal=cmatex,manuscript=article]{achemso}

\usepackage[version=3]{mhchem} 
\usepackage[T1]{fontenc}       

\author{Dmytro Dedovets}
\affiliation{Ceramic Synthesis \& Functionalization Lab, UMR3080 CNRS-Saint-Gobain, 84306 Cavaillon, France}
\author{C\'{e}cile Monteux}
\affiliation{SIMM, UMR 7615 CNRS-ESPCI-Universit\'e Pierre Et Marie Curie, ESPCI, Paris, France}
\author{Sylvain Deville}
\email{sylvain.deville@saint-gobain.com}
\affiliation{Ceramic Synthesis \& Functionalization Lab, UMR3080 CNRS-Saint-Gobain, 84306 Cavaillon, France}

\title{A temperature-controlled stage for laser scanning confocal microscopy and case studies in chemistry of materials}

\begin{document}

\begin{tocentry}

\includegraphics{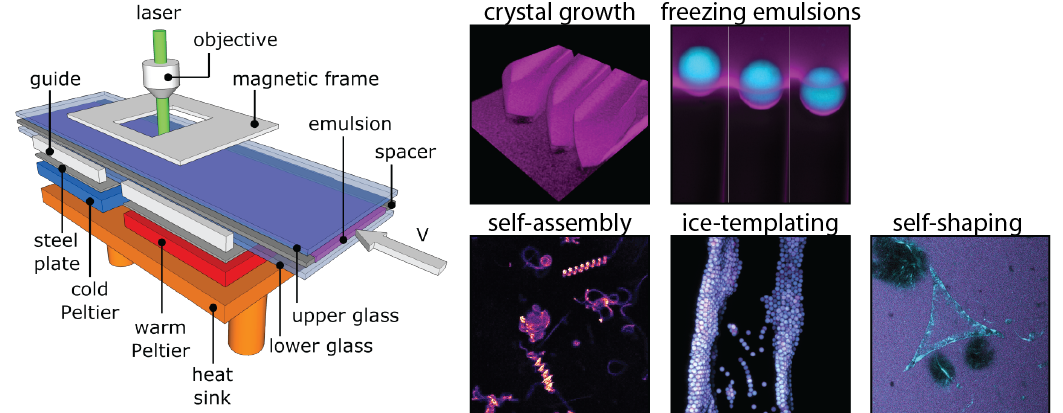}

\end{tocentry}

\begin{abstract}
If confocal microscopy is an ubiquitous tool in life science, its applications in chemistry and materials science are still, in comparison, very limited. Of particular interest in these domains is the use of confocal microscopy to investigate temperature-dependent phenomena such as self-assembly, diffusio- or thermophoresis, or crystal growth. Several hurdles must be solved to develop a temperature-controlled stage for laser scanning confocal microscopy, in particular regarding the influence of an elevated temperature gradient close to the microscope objective, which most people try very hard to avoid. Here we report the design of a temperature-controlled stage able to generate stable temperature gradients in both positive and negative temperature range and does not require use of liquid nitrogen. Our setup provides an excellent control of the temperature gradient, which can be coupled with a controlled displacement of the sample, making it useful in particular for a variety of solidification, chemistry, and interfacial problems. We illustrate the benefits of our setup with several case studies of interest in chemistry and materials science: the 3D real-time imaging of ice growth, the segregation of hard particles by growing crystals, the freezing behaviour of single emulsions, the self-shaping of oil droplets upon cooling, and the self-assembly of amphiphile molecules into helical structures. These results show how confocal microscopy coupled with a temperature-controlled stage is a welcome addition to the toolkit of chemists and materials scientists.

\end{abstract}
\maketitle

\section{Introduction}

Temperature gradients have an important impact on many natural and industrial, chemical, and physical processes ranging from geophysics and biology to metallurgy and materials synthesis. In Nature temperature gradients are responsible for the ice lensing of soils in cold regions and the metamorphism of snow~\cite{Wiese2017,Pinzer2012}, ice and the evolution of their interface~\cite{Hammonds2015}. Naturally occurring antifreeze proteins protect a variety of living species (plants, insects, fishes) from freezing. To evaluate their effectiveness and that of developed synthetic analogues, well established temperature gradients are required~\cite{Haleva2016}. In metallurgy, temperature gradients determine the morphology of solid/liquid interface~\cite{Clarke2017, Cai2016}, the lateral growth rate of the crystals~\cite{CHANG2012}, or the formation of intermetallic compounds at solid/liquid interfaces~\cite{Zhao2015}. The evolution of the mushy zone (partially solid/liquid region) in a fixed temperature gradient is also an important topic in nuclear safety\cite{Salloum-Abou-Jaoude2015}. Temperature gradient is one of the key parameter in the growth of monocrystals, which are essential components of optical systems~\cite{Natarajan2014} and microelectronics~\cite{Fujiwara2011}. Temperature gradients are used to create composition gradients for new types of materials such as plasmonic arrays~\cite{Ye2017}. In soft matter studies, temperature gradients are used to study self-assembly processes such as micelles formation~\cite{Jensen2013} or evolution of RNA-based self-assembled structures~\cite{Jaeger2005}. Temperature gradients are applied to study thermophoretic mobility of vesicles~\cite{Talbot2017}. Thermophoresis is used to manipulate bubbles and droplets\cite{Selva2010}. Temperature gradients were also used to study phase transitions in liquid crystals\cite{Balachandran2014} and later were applied to develop new temperature measurement techniques~\cite{Basson2012}. We could continue this list of examples for a long time.
Being able to use and control temperature gradients in experimental setups is therefore an problem with far-reaching implications. While sample heating is easily achieved by the use of resistive heaters, sample cooling typically relies on the use of liquid nitrogen which is, in some cases, inconvenient. In addition, because of thermal stability and condensation issues, most microscopists try very hard to avoid freezing temperatures and temperature gradients close to the microscope objectives. A few (costly) commercial temperature-controlled stages, cooled using liquid nitrogen, are available.  

Here we report the design of a temperature-controlled stage (and the corresponding samples) able to generate stable temperature gradients in both positive and negative temperature range and does not require the use of liquid nitrogen. We developed this stage to perform in situ laser scanning confocal microscopy, which is relevant to investigate a broad range of chemistry problems, as we will demonstrate in this paper. 

Confocal microscopy with rapid imaging capacities provides access to the 2D or 3D evolution of sample morphology with good time and space resolution. It allows simultaneous imaging of liquid phase, crystals and dispersed phase (particles, droplets, bubbles\ldots).

\section{Design of the stage}

\begin{figure}
\centering
\includegraphics[width=16cm]{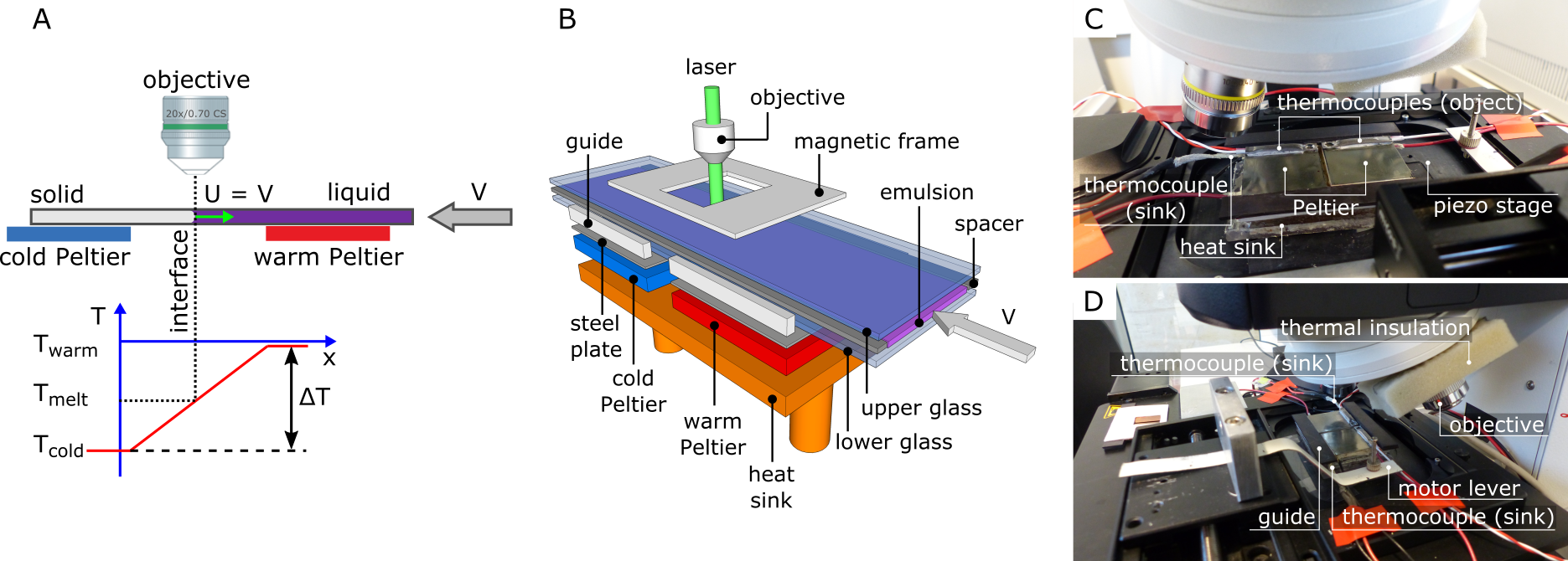}
\caption{Overview of the temperature gradient stage for in situ operations under the confocal microscope: (A) Operation schematics. A constant temperature gradient is achieved in the gap between the Peltier elements. If the sample is pushed at a velocity $V$, a solidification front moves at a velocity $U=V$. The solidification front is thus always in the centre of the observation windows, which is useful for solidification studies. (B) 3D representation. (C and D) Photographs of front and side views. The insulating foam that covers the sample was removed for the picture.}
\label{fig:figure1}
\end{figure}

We designed a temperature-controlled stage based on two Peltier elements (Fig.~\ref{fig:figure1}). The main advantage of this design is that it allows both heating and cooling. Therefore, temperature controlled experiments can be performed either in positive or negative temperature range, or in a mixed regime, providing elevated temperature gradients. 

For our stage, we selected two ET-127-10-13 Peltier cooler modules (Produced by Adaptive and purchased from RS Components, France). These modules have following parameters: dimensions = $30 \times 30 \times 3.6~mm$, $I_{max}$ = 3.9~A, $V_{max}$ = 15.7Vdc and maximum cooling power $Pc_{max}$ = 35.2~W. Dimensions were chosen based on the sample geometry and desired stage size. $I_{max}$ and $V_{max}$ determine the choice of power and controlling modules. The maximum cooling power defines the overall performance of the temperature-controlled stage and the setup requirements for the heat sink.

To control the Peltier cooling modules, we use TEC-1122 Dual Thermo Electric Cooling Temperature Controller from Meerstetter Engineering, Switzerland. This specialized TEC controller is able to independently drive two Peltier modules which generate an output current up to 10~A and driving voltage up to 21~V. The controller can handle Pt100, Pt1000, or NTC temperature probes. The temperature precision/stability of the controller is $<0.01^{\circ}$C. The temperature controller is paired to a laptop via USB. The TEC Service software (v2.50, provided with the controller) is used to monitor and control all the operating parameters of the temperature-controlled stage.

\begin{figure}
\centering
\includegraphics[width=8cm]{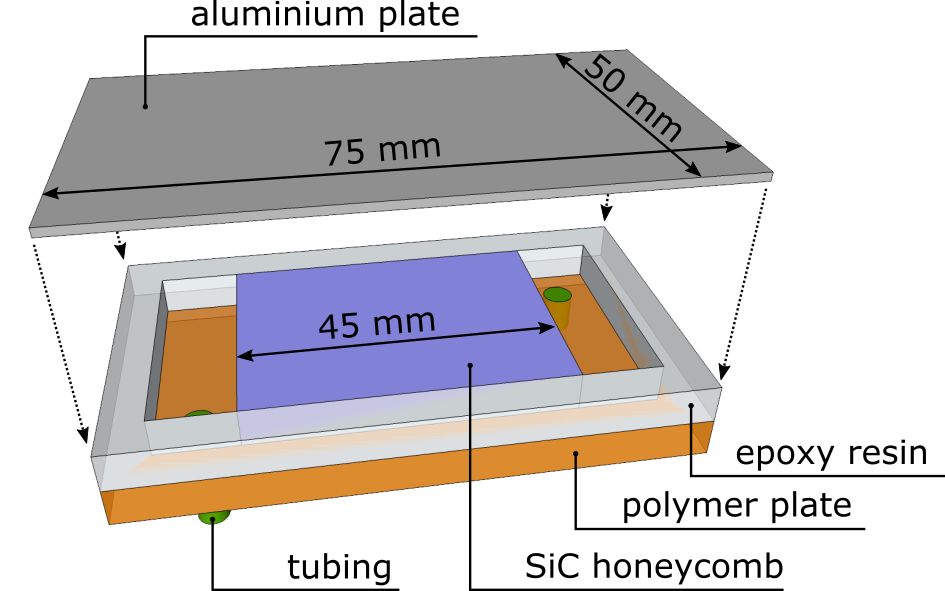}
\caption{Schematics of the heat sink.}
\label{fig:figure2}
\end{figure}

The design of an efficient heat sink is of uttermost importance to ensure the efficient operation of Peltier elements. The heat sink must withdraw and dissipate the heat generated on the warm side of the Peltier modules. The heat load is composed of the heat load of the object (amount of heat withdrew by Peltier element from the object) and heat load of the Peltier element due to losses. In our case, the maximum heat load can be as high as 96.4~W ($15.7~V \times 3.9~A + 35.2~W$) for each Peltier. Almost 200~W must thus be removed continuously during the operations. Based on these high heat loads, we design a home-made water-cooled heat sink (Fig.~\ref{fig:figure2}). A silicon carbide honeycomb (which provides a high surface exchange), in contact with the heat source, is used to transfer the heat to the coolant that circulates through the heat sink. The thermal conductivity of silicon carbide is sufficient to remove all the generated heat, while the chemical inertia of silicon carbide ensures that the heat sink does not degrade over time because of corrosion. The circulating water is cooled by MPC Minichiller (Model 3006.0015.99, Huber, Germany), which typically operates at $-5^{\circ}$C.

The temperature-controlled stage is thermally isolated with a polyurethane foam. A hole is cut in the foam to accommodate the objective of the microscope. This allows to decrease the amount of condensation on the top of the sample. 

An additional step is required to ensure a reliable contact with the Peltier elements and avoid undesirable z-translations of the sample during the operations. We thus attached thin (1~mm) steel plates on the top of each Peltier element. A magnetic frame is placed on the top of the sample to ensure that the sample remains in contact with the temperature-controlled stage during translation (Fig.~\ref{fig:figure1}B). This approach ensures a good thermal contact between the sample and the stage independently of sample thickness variations and guarantees that friction remains reasonably low to prevent stick-and-slip behaviour. Steel plates, in addition, help better homogenize the temperature field.

The temperature gradient stage is placed on the piezoelectric stage of the microscope. Because the piezoelectric stage needs to move freely during 3D acquisition, the weight placed on top of the stage cannot exceed $200~g$. It was therefore critical to come up with a lightweight design for the stage. The whole temperature-controlled stage weight is $160~g$, which ensures that 3D acquisition can be properly conducted.


\section{Price estimate} 

A breakdown of the cost of the elements used for the setup is given in Tab.~\ref{tbl:stage_price}. The most costly parts are the temperature controller and the water cooler (which most labs already have). For the later, we used an expensive (but very efficient) water cooler (4,000~Eur.), although a cheaper one would probably do the job as well. The overall cost is therefore in the 3,800--6,8000~Eur. range, which is much cheaper than the commercially available stages.  The cost does not include, of course, the time spent on designing, building, and testing the stage.

\begin{table}
  \begin{tabular}{lp{8cm}}
  Parts & Price (Eur.) \\
  \hline
  Temperature controller                  & 2,600\\
  Peltier elements ($\times 2$)           & 70\\
  Sink temperature sensors                & 6 \\
  Object temperature sensors ($\times 2$) & 104 \\
  Thermal paste                           & 25 \\
  Epoxy resin                             & 6\\
  Heat sink                               & 0 (spare parts) \\
  Water cooler                            & 1,000 to 4,000\\
  \hline
  Total                                   & 3,800 to 6,800\\
  \end{tabular}
  \caption{Price estimate of the different elements used to build the setup.}
  \label{tbl:stage_price}
\end{table}

\section{Sample preparation}

\begin{figure}
\centering
\includegraphics[width=8cm]{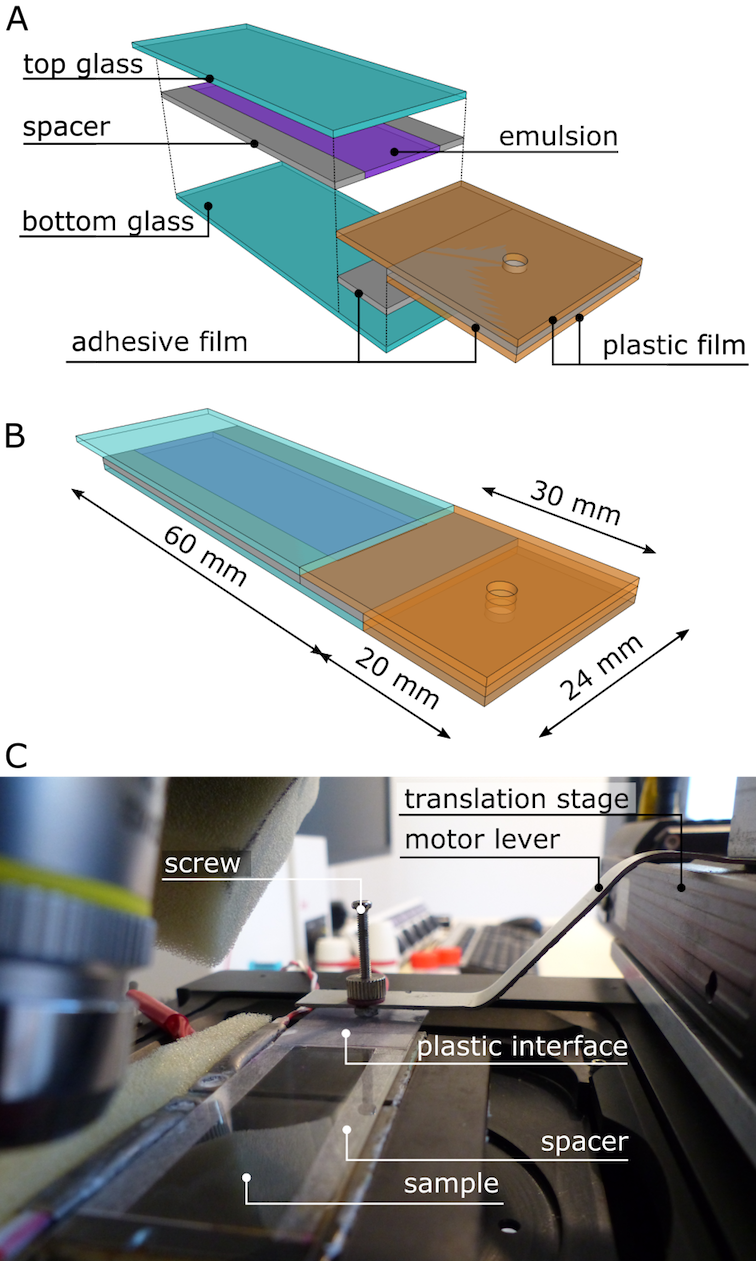}
\caption{Sample preparation (A and B). Photo of the assembled sample on the temperature-controlled stage (C).
\label{fig:figure3}}
\end{figure}

We design lightweight samples for solidification studies (see case studies below). Hele-Shaw cells are prepared by sandwiching the sample (solution, emulsion, suspension) between two microscope cover glass slides. The slides are connected together with two stripes of double-side sticky tape (Fig.~\ref{fig:figure3}A, B). The second stripe acts as a spacer and ensures a constant sample thickness. To reduce thermal mass of the sample cell we use cover glass slides (Menzel, $24 \times 60~mm$,  thickness 0.13--0.16~mm) instead of traditional microscope slides used in commercially available temperature-controlled stages. This solution decreases the thermal inertia in the system. However, it requires additional measures to ensure a constant contact of the sample with the stage during translation as we discussed above (magnetic frame). Samples were sealed with acrylic glue or nail polish. We observed no effect of possible solvent residues on the freezing behaviour of the samples.

For the experiments--solidification studies--where the sample is translated through a temperature gradient, we use Micos Pollux Drive stepper motor with VT-80 translation stage (PI, USA). A flexible connection of the sample to the motor was achieved via $0.1mm$ thick plastic strip. It is attached to the sample with double sided sticky tape and screwed to the motor lever (Fig.~\ref{fig:figure3}C). Such type of connection diminishes the risks of damages to the sample by bending loads arising from the lack of parallelism between the latter and the motor. It reduces as well heat transfer in the system.

For image acquisition, we use long working distance non-immersive objectives (Leica HC PL APO 20x/0.70 CS and 10x/0.40 CS2) to minimize the effect of the microscope thermal mass on the freezing process. These objectives have free working distances of 0.59~mm and 2.2~mm respectively. 

\section{Operation modes}

Several operation modes are possible, depending on the system investigated. In the case studies reported below, we used two main operating modes: (1) a fixed temperature gradient, no sample displacement and (2) a fixed temperature gradient with sample displacement. We use the first mode to investigate temperature-dependent phenomena such as self-assembly or the self-shaping of droplets. Because of the high temperature gradient (up to $20^\circ C/mm$) in a small observation window typically less than $1~mm$, it is possible to investigate a large variety of experimental conditions at once. Although we work at a constant temperature and temperature gradient, it is also possible to keep the temperature gradient constant while the temperature is increased or decreased, if needed. 

The second mode (sample displacement, fixed temperature gradient) is used for solidification studies. This mode has two main benefits. First, it allows to decouple the temperature gradient with the growth rate of the crystals, because they are independently controlled. These two parameters play a critical role on the solidification regimes and microstructure. The second benefit is a consequence of the fixed interface position in the observation frame. If we investigate the interactions of objects (particles, oil droplets) with the solidification front, we can observe and analyse hundreds or more of interaction events. A statistical analysis of the interactions can thus be achieved. We use this mode to investigate the 3D morphology of ice growth, the behaviour of hard particles in interaction with a solidification front, and the freezing of emulsions.

The typical operation procedure is as follows. The sample is prepared as described previously. The temperature of the stage is adjusted to $20^{\circ}$C. The sample is deposited on the freezing stage, attached to the stepper motor, and put-in-place with the magnetic frame. The stage is then thermally isolated with the polymer foam. The desired temperature gradient is established by setting the temperatures of Peltier elements. Finally--if needed--, the solidification front velocity is set by adjusting the speed of translation stage motor.

Image analysis was done with Fiji (ImageJ 1.51h)\cite{Schindelin2012}.

\section{Performances}

\begin{figure}
\centering
\includegraphics[width=8cm]{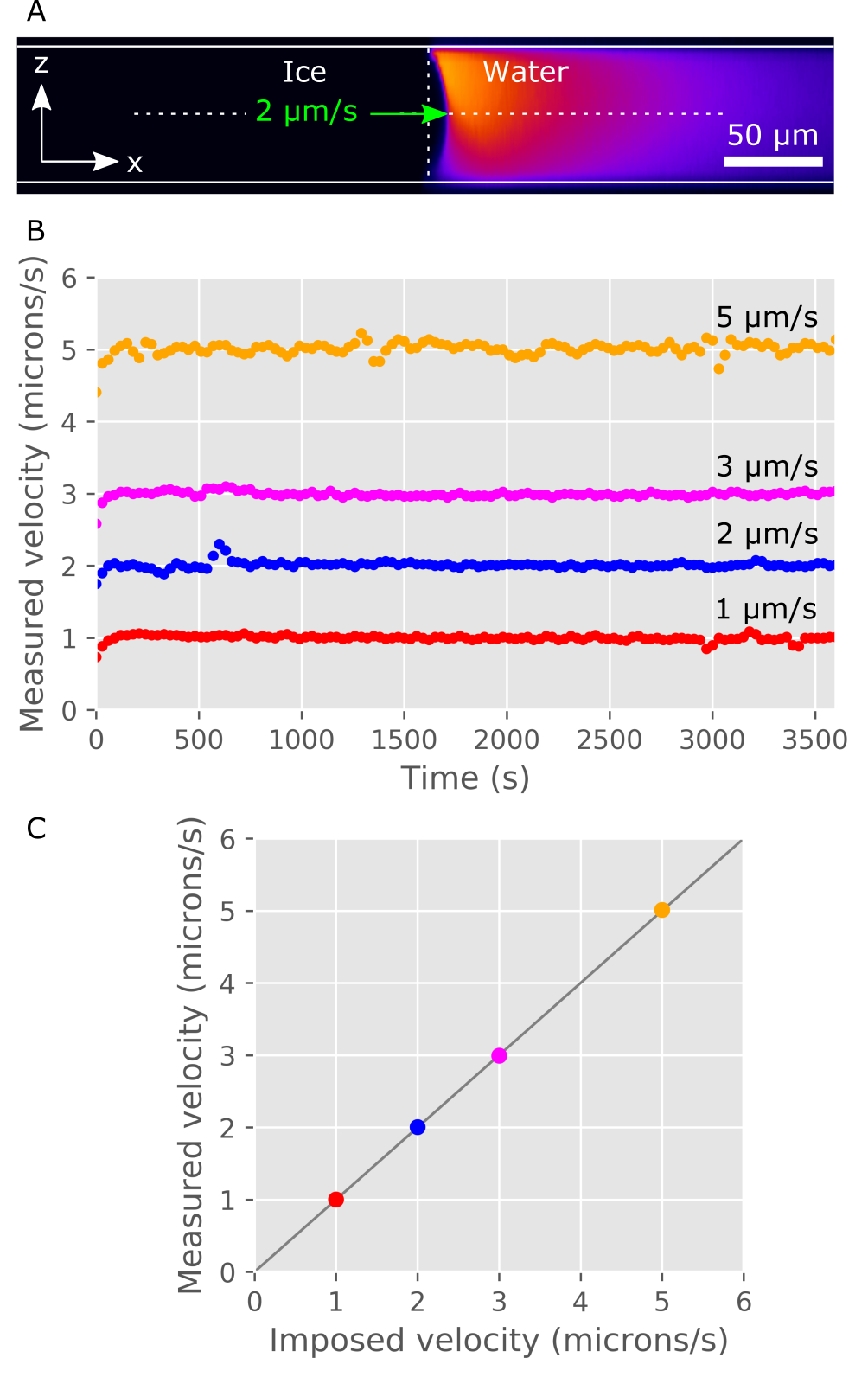}
\caption{(A) Cross-section of frozen $10^{-5}M$ dye solution along the temperature gradient of $10^{\circ} C/mm$. Sample displacement velocity $2\mu m/s$. (B) Measured solidification front velocity for several sample displacement velocities. (C) Measured solidification front velocity vs imposed one. The solid line represents a 1:1 dependence. Error bars do not exceed the size of the marker.
\label{fig:figure4}}
\end{figure}

The temperature of the Peltier elements in our setup can be varied from $-25^{\circ}$C to over $+90^{\circ}$C. The lower limit is determined by the choose of Peltier elements, efficiency of heat sink, temperature of cooling water, and thermal insulation of the stage. The upper limit is set by the sample evaporation and thermal stability of some elements of the stage. The maximal temperature gradient that can be established depends on the temperature of each Peltier. In our experiments, we apply temperature gradients in the 5--20$^{\circ}C/mm$ range. The design of the stage, combined with the small sample thickness, ensure that the vertical temperature gradient through the sample is negligible. Symmetrical ice crystals are formed (Fig.~\ref{fig:figure4}A). 

Pt100 thermocouples provide temperature readings with a precision of $\pm 0.05\%$ while the temperature controller ensures a temperature precision of $0.01^{\circ} C$. The actual precision of the temperature control within the sample is, however, lower due to the sample displacement and therefore to the dynamic character of the thermal contact with the temperature-controlled stage. Position of solid/liquid interface for the temperature gradient of $10^{\circ}C/mm$ and several sample displacement velocities are shown in Fig.~\ref{fig:figure4}B. The interface velocity stabilizes within a minute after the beginning of the sample translation. The small fluctuations observed, most likely, arise from the variations in the thermal contact between the sample and the freezing stage.

We find a very good agreement between the imposed velocity (speed of stepper motor) and the measured velocity of the solidification front (Fig.~\ref{fig:figure4}C). The VT-80 translation stage can adjust the sample displacement velocity from $0.5\mu m/s$ to several $mm/s$. The useful upper limit is nevertheless mainly determined by the acquisition performance of the microscope and specificity of the experiment.

\section{Design of the solutions/emulsions}

We choose a set of three oils combining a melting temperature significantly lower than that of the water, a low solubility in water and compatibility with standard surfactants--see Tab.~\ref{tab:oil}. For the most of the studies we used propyl benzoate as an oil due to the good density matching with water. 

\begin{table}
\begin{tabular}{lcccc}
Oil             & Melting point   & Solubility ($g/100g$) & Density ($g/cm^3$) & CAS number \\
\hline
Anisole         & $-37^\circ C$   & 0.100                 & 0.995              & 100-66-3   \\
Octyl acetate   & $-38.5^\circ C$ & 0.018                 & 0.870              & 112-14-1   \\
Propyl benzoate & $-51.6^\circ C$ & 0.035                 & 1.023              & 2315-68-6  \\
\end{tabular}
\caption{Properties of suitable oil candidates to prepare the emulsions for freezing experiments.}
\label{tab:oil}
\end{table}

\begin{table}
\begin{tabular}{lccc}
Dye              & Excitation wavelength (nm) & Emission wavelength (nm) & CAS number \\
\hline
Sulforhodamine B & 565                        & 586                      & 3520-42-1\\
$BODIPY^*$       & 493                        & 504                      & 121207-31-6\\
\end{tabular}
\caption{Properties of dyes used to prepare the emulsions for freezing experiments. The emission and excitation wavelengths were measured respectively in water and in methanol. $^*$ Difluoro{2-[1-(3,5-dimethyl-2H-pyrrol-2-ylidene-N)ethyl]-3,5-dimethyl-1H-pyrrolato-N}boron. This is a more hydrophobic version of the BODIPY fluorescent dye. }
\label{tab:dye}
\end{table}

To dye the aqueous phase we use Sulforhodamine B. To avoid fluorescence self-quenching which was reported above $2\times10^{-4} M$\cite{Arbeloa1989} we use $10^{-5} M$ solution in combination with 1\% - 5\% laser power (MAX power - $10~mW$) for all the experiments. For oil staining we use Difluoro{2-[1-(3,5-dimethyl-2H-pyrrol-2-ylidene-N)ethyl]-3,5-dimethyl-1H-pyrrolato-N}boron. This is a more hydrophobic version of the BODIPY fluorescent dye. It has good solubility in all the oils we studied. A concentration of $10^{-4} M$ is sufficient to obtain clear imaging at 1\% laser power. We observe no significant changes in the fluorescence intensity upon cooling.

\section{Case studies}

We illustrate here the benefits of our setup with a variety of case studies of interest in chemistry in particular, and more generally in materials science: 3D real-time imaging of ice growth, the segregation of hard particles by growing crystals, the freezing behavior of single emulsions, the self-shaping of oil droplets upon cooling, and the self-assembly of amphiphiles into helical structures.

\subsection{3D, real-time imaging of ice growth}

An advantage of confocal microscopy in comparison to other characterization techniques is ability to perform \textit{in situ} 3D imaging of the sample without introducing observation artifacts. A number of techniques have been used to investigate crystal growth, and each one of them has its own benefits and limitations. Optical microscopy is easy to perform, does not induce observation artifacts (low energy of the light beam), but does not provide a 3D representation of the crystals. The only technique able to provide in situ 3D representation of crystal growth is X-ray tomography. However, the effects of interaction between the highly energetic beam and the ice crystals are still problematic~\cite{Deville2013}. Confocal microscopy provides thus a complementary alternative to X-ray imaging.

We previously used confocal microscopy to image crystal growth \textit{in situ}~\cite{Marcellini2016}. However, the previous setup provided little to no control over the cooling rate and temperature gradient, and the solidification front was moving through the observation window, making the operation difficult. With the current temperature-controlled stage, we have a complete control of the growth velocity, growth direction, and temperature gradient. It is thus much easier to investigate in a reliable and reproducible manner.

\begin{figure}
\centering
\includegraphics[width=8cm]{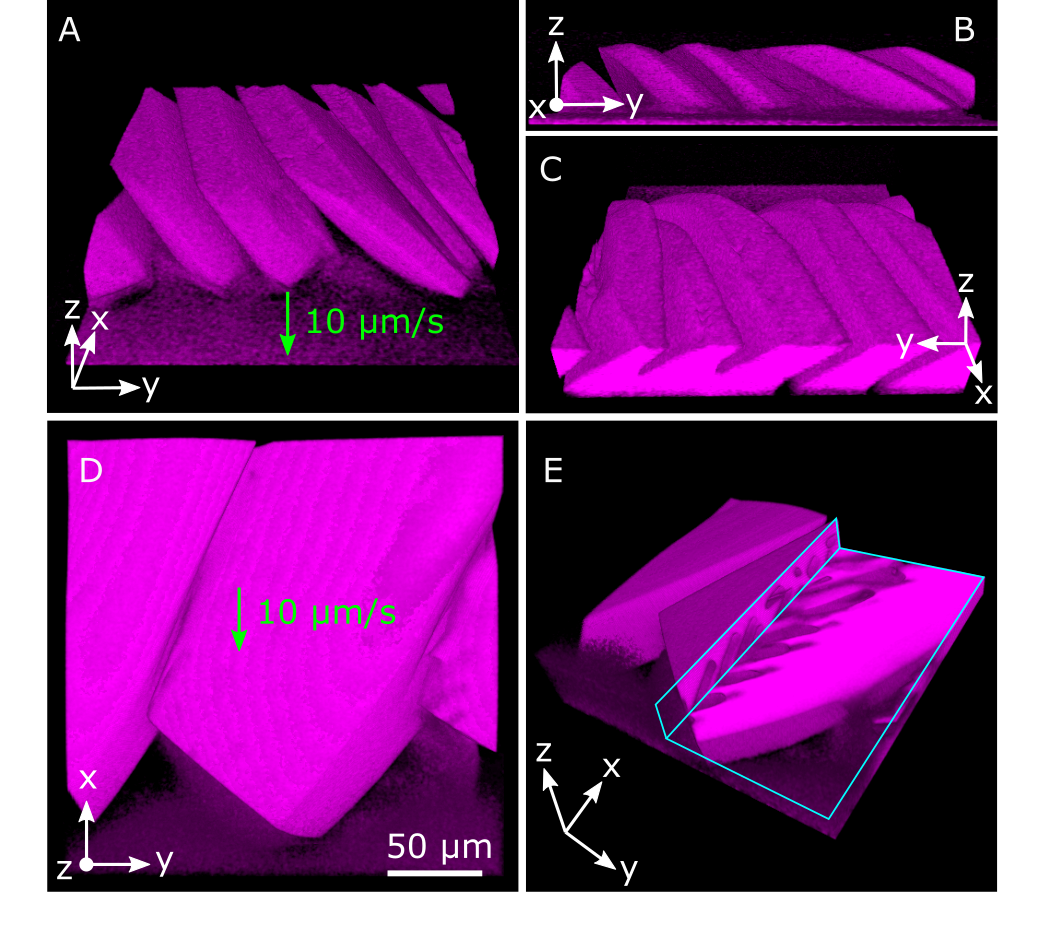}
\caption{3D reconstruction of ice crystals formed in $200mM$ zirconium acetate solution at temperature gradient of: (A-C) $15^{\circ}C/mm$ and (D-E) $5^{\circ}C/mm$. The green arrows indicate the growth direction and rate of crystal growth. Cyan rectangles indicate planes along which 3D reconstruction was sectioned. Reconstructed volume is $291 \times 291 \times 58 \mu m^3$ for the sample (A-C) and $291 \times 291 \times 67 \mu m^3$ for the sample (D-E).
\label{fig:figure5}}
\end{figure}

This is particularly useful when one of the main crystallographic axes of the sample does not coincide with the orientation of the focal plane. One such example is the growth of tilted crystals during the solidification of $200mM$ solution of ice shaping compound - zirconium acetate (Fig.~\ref{fig:figure5}A-C). Classical 2D imaging in this case will not provide the complete information on the sample morphology. Volume information acquired in such a case with AFM may be misinterpreted as well.

The importance of 3D imaging becomes even more prominent if we reduce the temperature gradient in this system from $15^{\circ}C/mm$ to $5^{\circ}C/mm$ . This leads to the formation of internal porosity with complex geometry (Fig.~\ref{fig:figure5}E) while in-plane view provides no evidences of such change in the ice shaping behavior of zirconium acetate (Fig.~\ref{fig:figure5}D). However, this observation of the internal porous structure of the crystals help us rationalize the previously reported porous morphology of ice-templated materials, obtained under similar conditions~\cite{Marcellini2016}. 

\subsection{Segregation of hard particles by growing crystals}

The interaction of particles with a solidification front has been a central feature of interest in solidification studies for the past 40 years. Because of its many occurrences in different domains, from geophysics (soils freezing) to metallurgy (particle-reinforced alloys), food engineering, or ice-templating of porous materials~\cite{Deville2017}, a lot of attention has been paid to investigate the interactions of particles with growing crystals and their outcome. Of particular interest are the solidification patterns~\cite{Watanabe2000}, the development of secondary phases or premelted films around entrapped particles~\cite{Dash1999}, and the solidification in confined environments~\cite{HansenGoos2010,Coasne2013}. 

Further progress in our understanding of these phenomena requires \textit{in situ} observation of the solidification processes, at the spatial resolution of particles. In addition, being able to discriminate the solid and liquid phases during solidification is critical. Confocal microscopy, with point by point illumination of the sample and rejection of out of focus light, provides a way to overcome many of the problems in conventional microscopy caused by multiple scattering of objects which are out of focus, and prevents imaging deep within a sample. A popular technique in solidification studies is X-ray tomography. However, in addition to the beam-induced artefacts mentioned above, the current spatial resolution does not allow imaging individual particle and their dynamics.

Here we use our temperature-controlled stage to investigate the freezing of hard particles suspensions in conditions similar to that used in materials processing routes based on freezing~\cite{Deville2017g}. The 1~vol.\% particle suspension was frozen at $20\mu m/s$, in a $5^{\circ}C/mm$ temperature gradient.

\begin{figure}
\centering
\includegraphics[width=8cm]{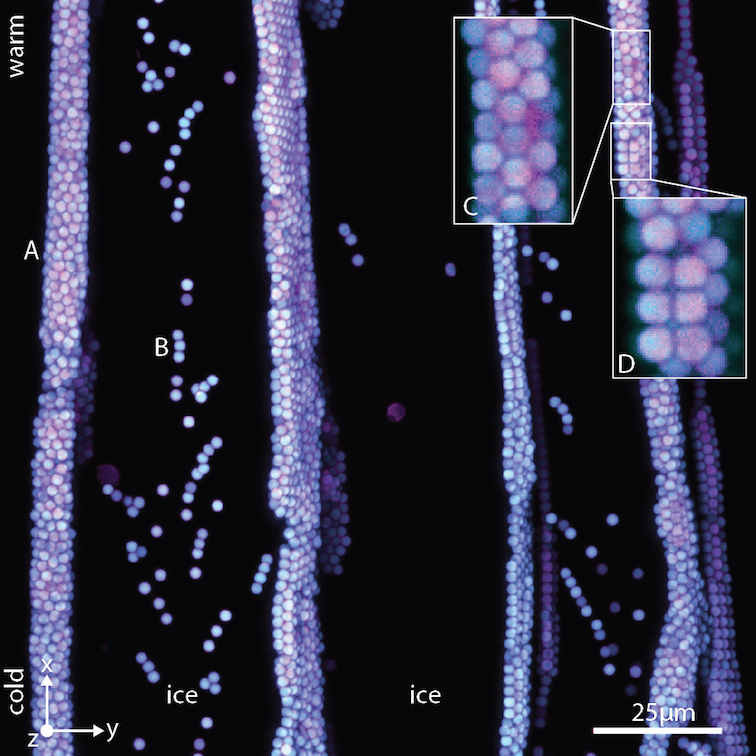}
\caption{Accumulation of solid particles between growing ice crystals during directional solidification. Crystals grew from the bottom (cold) to the top (warm). Most particles have been segregated and concentrated between the crystals. A few isolated particles have nevertheless been engulfed by growing crystals. Water is still liquid in most of the inter-particle spaces. Temperature gradient: $5^{\circ}C/mm$, growth rate $20\mu m/s$. 
\label{fig:figure6}}
\end{figure}

A 3D view of the frozen sample is shown in Fig.~\ref{fig:figure6}. The Sulforhodamine B dye is rejected by the ice crystals, the black regions of the image corresponds thus to the ice crystals. The rapid growth rate of the crystals (20 $\mu m/s$) induce lamellar solidification patterns typically encountered in ice-templating studies~\cite{Deville2007a}. The particles are segregated between the lamellar ice crystals (region A), where a nearly dense packing can be observed. We can also notice that not all particles are segregated by the crystals: a few isolated particles (region B) are entrapped within the ice crystals. This means that a mixed regime can exist during solidification, where most--but not all-- particles are segregated by the solidification front. The critical velocity above which particles are engulfed by the front ~\cite{Rempel2001a} directly depends on the particle size. However, here we use particles with a very narrow size distribution. Therefore the mixed regime cannot be explained by a particle size distribution effect, where large particles would be engulfed and smaller one repelled and concentrated between adjacent crystals.

Higher magnification observations (insets C and D in Fig.~\ref{fig:figure6}) enable us to identify the inter-particle regions where ice is already present. In inset C, a dense packing of particles can be seen. The pore size between the particles is thus small, and we can see on the image that the water is still liquid, as the Sulforhodamine B signal is strong. In inset D, however, a cubic packing with larger pores is observed. The Sulforhodamine B signal is much weaker, revealing thus that ice has already penetrated into the pores. This behavior can be explained by the Gibbs-Thomson freezing point depression: the freezing point decreases with pore size~\cite{Liu2003a}. Hence pore ice is formed first in the largest pores. With confocal microscopy, we can thus follow the progression of freezing within the packing of segregated particles. No other technique can provide such observations today.

The perspectives in the solidification studies in the presence of particles are therefore exciting. Not only can we image ice, liquid, and particles, but the temperature-controlled stage also provides a decoupled control of the temperature gradient and the ice growth rate, which are the main parameters that control the solidification patterns and segregation behaviours. In addition, we can perform such observations dynamically, and should thus be able to investigate the dynamics of particle segregation and their coupling with crystal growth.

\subsection{Freezing emulsions}

Emulsions are another interesting system to freeze, for several reasons. The first one is, of course, to better understand the freeze/thaw stability of emulsions, which are of interest in many chemistry~\cite{Lin2008}, pharmaceutical~\cite{Bogdan2016}, and food engineering~\cite{Ghosh2008,Degner2014,Katsuki2017} applications. The second motivation are freezing routes using emulsions, which can yield porous materials~\cite{Chatterjee2016} or capsules~\cite{Khapli2009}. The third reason, with far-reaching implications, is the use of model mono-dispersed emulsions to investigate the interaction of objects with a solidification front~\cite{Dedovets2017}. The later is a phenomenon encountered in applications as diverse as the freezing of soils in cold regions, the cryopreservation of cells, the solidification of particle-reinforced alloys, or the removal of pollutants by directional freezing~\cite{Deville2017b}. Because we can easily control their size, composition, and surface chemistry, oil droplets are interesting objects to play with in this case. They may offer a particularly interesting analogy to the behaviour of soft objects such as reproductive or red blood cells.

\begin{figure}
\centering
\includegraphics[width=8cm]{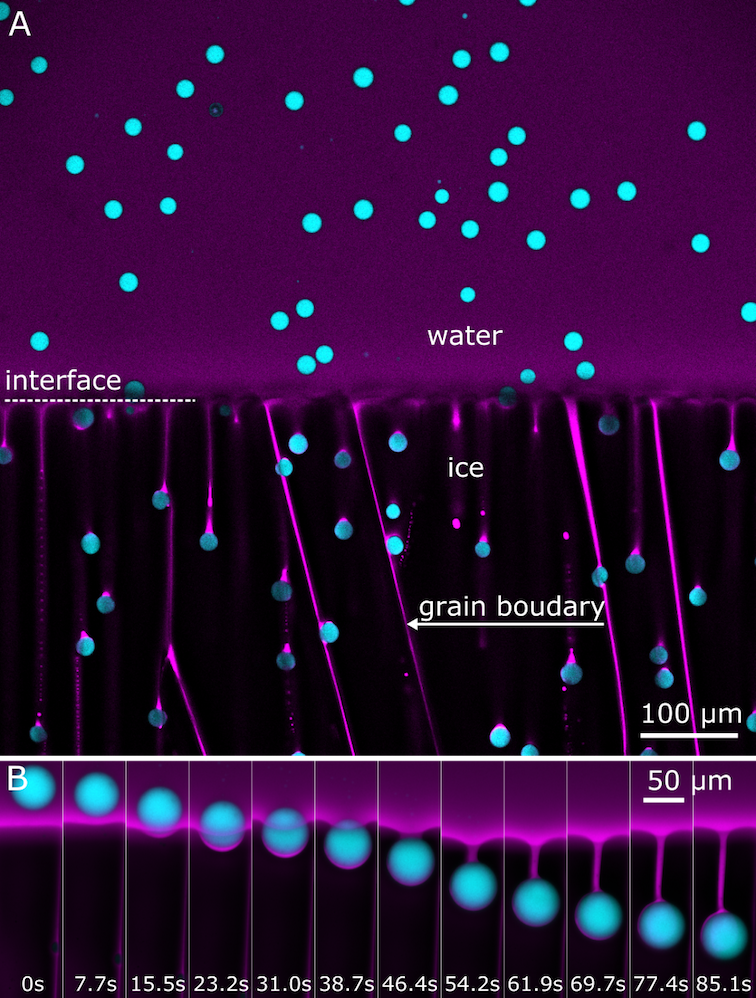}
\caption{Directional solidification of oil-in-water emulsion: (A) General view of the sample. (B) Time-lapse sequence of the oil droplet encapsulation by the solidification front.
\label{fig:figure7}}
\end{figure}

Before moving to systems closer to that of applications, it is better to start the investigations with a model system. Here we prepare an oil-in-water emulsion using microfluidics. The volume fraction of oil is low enough so that droplets do not interact with each other during freezing. We can thus investigate in details the interaction of isolated droplets with the front. A typical image is shown in Fig.~\ref{fig:figure7}. The ice rejects the dye from water and appears thus in black. The liquid regions fluoresce (magenta). A second dye, incorporated in the oil droplets (cyan), enables us to identify them. We can thus simultaneously image the liquid, the ice, and the droplets, in 2D or 3D. 

With rapid imaging capacities, we can capture in 3D the interaction of an oil droplet with the solidification front. A typical time-lapse, 3D reconstruction is shown in Fig.~\ref{fig:figure7}B. Here, we combined all the $z$ slices (z-projection). The result is equivalent to an optical microscopy image with an extended depth of view and limited light scattering. We can thus clearly follow the movement of the droplet, the deformation of the solid/liquid interface, as well as the formation of a thin liquid film between the droplet and the ice. This thin film is preserved when the droplet is completely engulfed by the ice. In the example shown here, the thin liquid film extends towards the solidification front, creating thus a defect (grain boundary) in the solidification microstructure. Because the solidification front is at a constant position in the observation frame, we can gather statistics about the droplet behaviour, unlike previous studies of particle/front interactions where only a few interactions events were imaged and analysed. Hundreds of interactions events can be analysed if the experiment is run for long enough~\cite{Dedovets2017}.

The current setup offers  a uniquely-controlled platform to investigate the freeze/thaw stability of emulsions, with 3D, in situ, multiphase imaging capacities, providing statistical informations about the interactions of soft objects with a solidification front. It may thus become a valuable tool in the study of such systems.

\subsection{Self-shaping of oil droplets upon cooling}

Bottom-up approach becoming an attractive energy and material efficient alternative to the classical top-down approach in the materials synthesis. A great variety of organic~\cite{Luo2012,Busseron2013,Hu2014}, inorganic~\cite{Yang2011} and composite materials~\cite{Ariga2008} was obtained via the bottom-up approach by using polymers, small synthetic and biological molecules, amphiphiles, graphene and other components as building blocks. Of particular interest of organic self-assembly is that many of such systems, including DNA~\cite{Rogers2016}, peptides~\cite{Wang2016} and amphiphiles~\cite{Sorrenti2013, Liu2013}, exhibit temperature metamorphism - i.e. different morphologies may be obtained by changing the temperatures or cooling rate. While being able to provide multiplicity of morphologies, self-assembling systems can be difficult to study when both thermodynamics and kinetics are involve in the structure formation~\cite{Wang2015}.

\begin{figure}
\centering
\includegraphics[width=16cm]{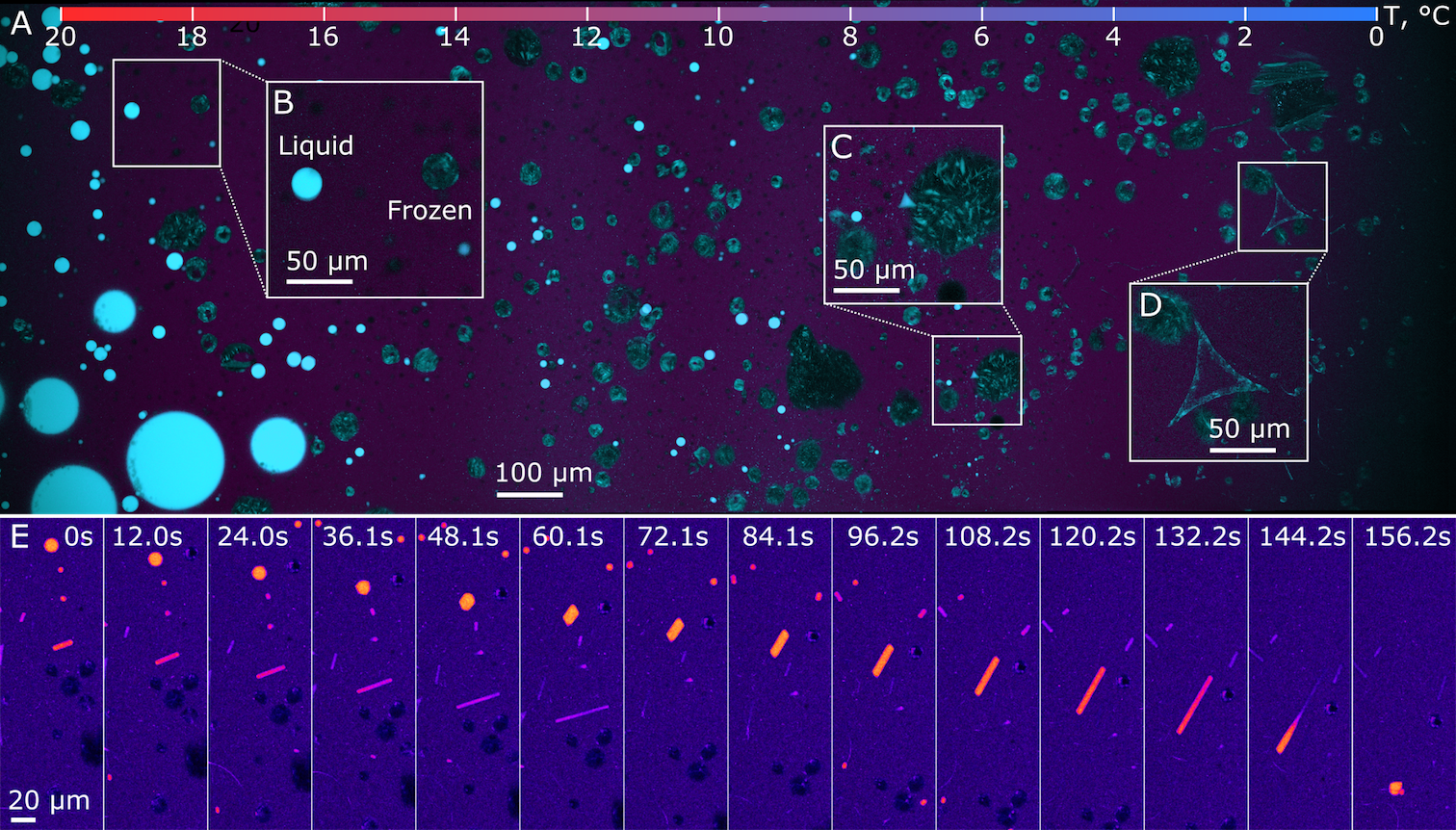}
\caption{Hexadecane in water (contains 1 wt\% of Brij 58 surfactant) emulsion in the temperature gradient of $10^{\circ}C/mm$: General view (A). Zoom-in on the different morphological features of the sample (B-D). Time-lapse sequence of the evolution of the sample translated through the temperature gradient of $10^{\circ}C/mm$ with the velocity of 1~$\mu m/s$. Different colour LUTs are applied to the images A--D and E for better visibility.
\label{fig:figure9}}
\end{figure}

Here we present an example based on the work of Denkov \textit{et al.}~\cite{Denkov2015}. Hexadecane droplets in water (containing appropriate surfactant) spontaneously change their morphology several times, generating series of complex regular shapes owing to the internal phase-transition processes (Fig.~\ref{fig:figure9}). The sample consists of 5\% v/v of hexadecane in water (containing 1 wt\% of Brij 58 surfactant). The emulsion is produced by hand shaking for 30 seconds. The sample is introduced at $20^{\circ}C$ and then one side was cooled to $0^{\circ}C$ with a cooling rate of $0.1^{\circ}C/s$. 

The confocal image is shown in Fig.~\ref{fig:figure9}A. At $18^{\circ}C$ (melting point of hexadecane), the oil freezes, resulting in the change in the droplet colour (insert B in Fig.~\ref{fig:figure9}). A variety of frozen droplets with irregular shapes including triangles (Fig.~\ref{fig:figure9}C and D), hexagons, and prisms (Fig.~\ref{fig:figure9}E) can be observed below the melting point.

Fig.~\ref{fig:figure9}E shows the time-lapse sequence for the sample translated with a velocity of 1~$\mu m/s$ over a temperature gradient of $10^{\circ}C/mm$. This corresponds to the cooling rate of $0.01^{\circ}C/s$. In total agreement with the work of Denkov \textit{et al.} under slow cooling spherical oil droplet transforms consecutively into hexagonal platelet, tetragonal platelet, platelet with high aspect ratio and, eventually, thin fibre. The advantage of the approach proposed here over the classical cooling experiment is that a large amount of statistical data on the droplets metamorphism can be accumulated within single experiment.

\subsection{Self-assembly of amphiphiles into helical structures}

Chiral assemblies such as twisted ribbons, helices and tubes are commonly encountered in Nature, with the double helix of DNA being the most well-known. Such structures have attracted extensive research interest over the last decade owing to their importance for the understanding of biological phenomena and use in nanotechnology~\cite{Yang2011} and functional materials synthesis~\cite{Luo2012, Busseron2013}.

Chiral structures are typically formed upon the cooling of the sample below the phase transition temperature~\cite{Ziserman2011}. Determining the effect of the temperature on the formation kinetics and morphology of aggregates can be, however, a time consuming task. Here we present a fast screening of the temperature effect on the amphiphile self-assembly.

The following experiment is based on the work of Aime \textit{et al.}~\cite{Aime2007,Aime2009}. The sample contains $20mM$ didodecyldimethylammonium bromide, $20mM$ guanosine 5'-monophosphate (GMP) disodium salt, and acetic acid to adjust the pH to 5.9. We first solubilize the sample at $50^{\circ}C$ and then introduced it in a temperature gradient of $10^{\circ}C/mm$ (between $20^{\circ}C$ and $40^{\circ}C$).  

\begin{figure}
\centering
\includegraphics[width=16cm]{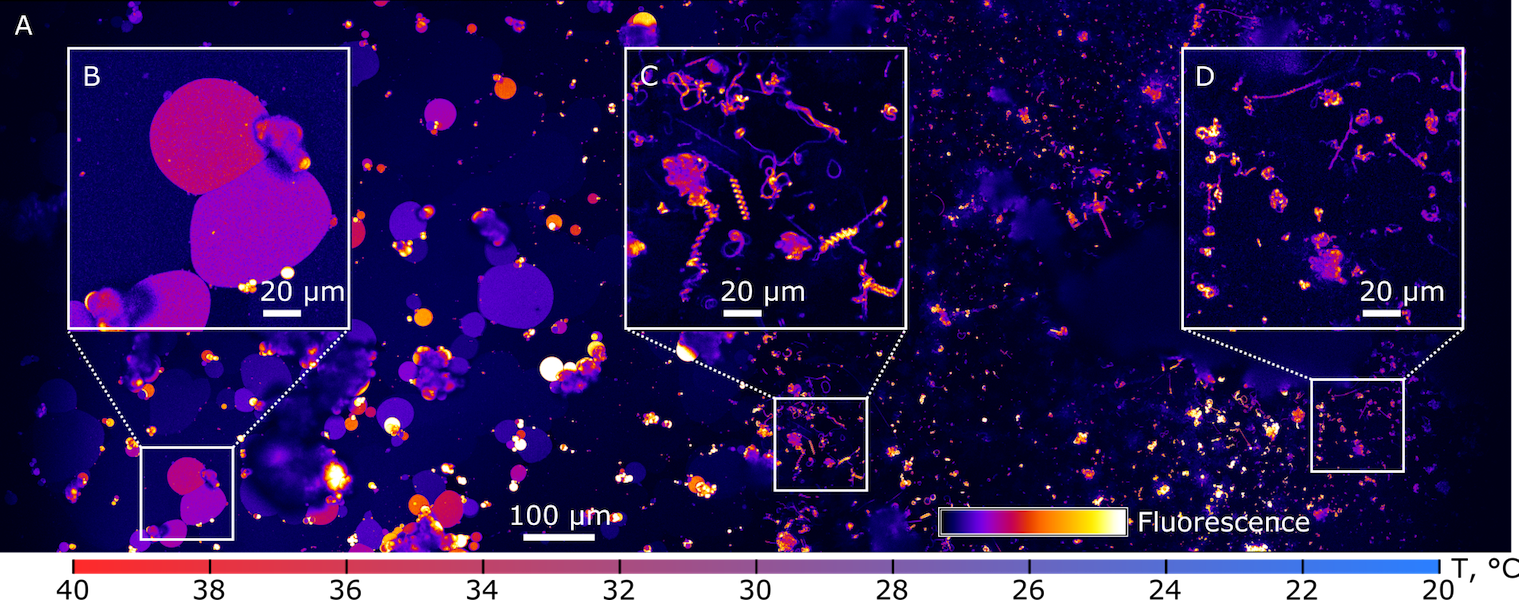}
\caption{Organic self-assembly in linear temperature gradient of $10^{\circ}C/mm$: (A) overview; (B) zone above the Krafft temperature (KT): melted sample; (C) zone below the KT: helical structures, fast kinetics; (D) zone below the KT: helical structures, slow kinetics. Image recorded 4 hours after the beginning of experiment.
\label{fig:figure10}}
\end{figure}

Fig.~\ref{fig:figure10}A shows the sample morphology after four hours of ageing. Sulforhodamine B fluorescent dye forms strong ion pair with amphiphile, which ensures that in our experiment we do visualize only the latter. The reported Krafft temperature of didodecyldimethylammonium guanosine 5'-monophosphate complex is $35^{\circ}C$ (for $3 mM$ solution). Above this temperature amphiphile is in the melted state. It forms circular domains at the surface of the bottom glass slide (Fig.~\ref{fig:figure10}B). Fluorescence intensity differs between the domains which can result from the thickness variation (multi-layered structure). Below the Krafft temperature, the amphiphile/GMP complex precipitates from the solution and self-assembles into micrometer-size helical structures (Fig.~\ref{fig:figure10}C). Interestingly, even close to the phase transition temperature, we do not observe any helix orientation along the temperature gradient. This suggests that helical structures are not formed through the condensation of monomers from solution, but rather via the structural reorganization of existing precipitate. Decreasing the temperature results in the formation of smaller helical structures evincing slower self-assembly kinetics (Fig.~\ref{fig:figure10}D).

\section{Conclusions and perspectives}

We presented here a versatile design of a stage for temperature controlled-experiments under the confocal microscope, including experiments with elevated temperature gradients (up to $20^\circ C/mm$ here). The variety of examples reported here, from crystal growth to the freezing stability of emulsions emulsions, the self-assembly of amphiphiles into helical structures or the self-shaping of oil droplets, illustrate the potential of this approach in chemistry and materials science. If confocal microscopy is an ubiquitous tool in life sciences, very little has been comparatively shown in chemistry and materials science. We hope that this study and setup design will inspire other chemists to enter this playground. While we used only two detectors in our setup, most confocal microscope today can accommodate up to 5 detectors. More complex multiphase imaging can thus be envisioned if needed.

\begin{acknowledgement}
The research leading to these results has received funding from the European Research Council under the European Union's Seventh Framework Programme (FP7/2007-2013) / ERC grant agreement 278004 (project FreeCo).  We thank Tom Kodger for providing the particles used in one of the examples shown here.
\end{acknowledgement}

\bibliography{freezing_stage}

\providecommand{\latin}[1]{#1}
\makeatletter
\providecommand{\doi}
  {\begingroup\let\do\@makeother\dospecials
  \catcode`\{=1 \catcode`\}=2\doi@aux}
\providecommand{\doi@aux}[1]{\endgroup\texttt{#1}}
\makeatother
\providecommand*\mcitethebibliography{\thebibliography}
\csname @ifundefined\endcsname{endmcitethebibliography}
  {\let\endmcitethebibliography\endthebibliography}{}
\begin{mcitethebibliography}{55}
\providecommand*\natexlab[1]{#1}
\providecommand*\mciteSetBstSublistMode[1]{}
\providecommand*\mciteSetBstMaxWidthForm[2]{}
\providecommand*\mciteBstWouldAddEndPuncttrue
  {\def\EndOfBibitem{\unskip.}}
\providecommand*\mciteBstWouldAddEndPunctfalse
  {\let\EndOfBibitem\relax}
\providecommand*\mciteSetBstMidEndSepPunct[3]{}
\providecommand*\mciteSetBstSublistLabelBeginEnd[3]{}
\providecommand*\EndOfBibitem{}
\mciteSetBstSublistMode{f}
\mciteSetBstMaxWidthForm{subitem}{(\alph{mcitesubitemcount})}
\mciteSetBstSublistLabelBeginEnd
  {\mcitemaxwidthsubitemform\space}
  {\relax}
  {\relax}

\bibitem[Wiese and Schneebeli(2017)Wiese, and Schneebeli]{Wiese2017}
Wiese,~M.; Schneebeli,~M. {Early-stage interaction between settlement and
  temperature-gradient metamorphism}. \emph{Journal of Glaciology}
  \textbf{2017}, \emph{63}, 652--662\relax
\mciteBstWouldAddEndPuncttrue
\mciteSetBstMidEndSepPunct{\mcitedefaultmidpunct}
{\mcitedefaultendpunct}{\mcitedefaultseppunct}\relax
\EndOfBibitem
\bibitem[Pinzer \latin{et~al.}(2012)Pinzer, Schneebeli, and
  Kaempfer]{Pinzer2012}
Pinzer,~B.~R.; Schneebeli,~M.; Kaempfer,~T.~U. {Vapor flux and
  recrystallization during dry snow metamorphism under a steady temperature
  gradient as observed by time-lapse micro-tomography}. \emph{Cryosphere}
  \textbf{2012}, \emph{6}, 1141--1155\relax
\mciteBstWouldAddEndPuncttrue
\mciteSetBstMidEndSepPunct{\mcitedefaultmidpunct}
{\mcitedefaultendpunct}{\mcitedefaultseppunct}\relax
\EndOfBibitem
\bibitem[Hammonds \latin{et~al.}(2015)Hammonds, Lieb-Lappen, Baker, and
  Wang]{Hammonds2015}
Hammonds,~K.; Lieb-Lappen,~R.; Baker,~I.; Wang,~X. {Investigating the
  thermophysical properties of the ice-snow interface under a controlled
  temperature gradient. Part I: Experiments {\&} Observations.} \emph{Cold
  Regions Science and Technology} \textbf{2015}, \emph{120}, 157--167\relax
\mciteBstWouldAddEndPuncttrue
\mciteSetBstMidEndSepPunct{\mcitedefaultmidpunct}
{\mcitedefaultendpunct}{\mcitedefaultseppunct}\relax
\EndOfBibitem
\bibitem[Haleva \latin{et~al.}(2016)Haleva, Celik, Bar-Dolev, Pertaya-Braun,
  Kaner, Davies, and Braslavsky]{Haleva2016}
Haleva,~L.; Celik,~Y.; Bar-Dolev,~M.; Pertaya-Braun,~N.; Kaner,~A.;
  Davies,~P.~L.; Braslavsky,~I. {Microfluidic Cold-Finger Device for the
  Investigation of Ice-Binding Proteins}. \emph{Biophysical Journal}
  \textbf{2016}, \emph{111}, 1143--1150\relax
\mciteBstWouldAddEndPuncttrue
\mciteSetBstMidEndSepPunct{\mcitedefaultmidpunct}
{\mcitedefaultendpunct}{\mcitedefaultseppunct}\relax
\EndOfBibitem
\bibitem[Clarke \latin{et~al.}(2017)Clarke, Tourret, Song, Imhoff, Gibbs,
  Gibbs, Fezzaa, and Karma]{Clarke2017}
Clarke,~A.~J.; Tourret,~D.; Song,~Y.; Imhoff,~S.~D.; Gibbs,~P.~J.;
  Gibbs,~J.~W.; Fezzaa,~K.; Karma,~A. {Microstructure selection in thin-sample
  directional solidification of an Al-Cu alloy: In situ X-ray imaging and
  phase-field simulations}. \emph{Acta Materialia} \textbf{2017}, \emph{129},
  203--216\relax
\mciteBstWouldAddEndPuncttrue
\mciteSetBstMidEndSepPunct{\mcitedefaultmidpunct}
{\mcitedefaultendpunct}{\mcitedefaultseppunct}\relax
\EndOfBibitem
\bibitem[Cai \latin{et~al.}(2016)Cai, Wang, Kao, Pericleous, Phillion, Atwood,
  and Lee]{Cai2016}
Cai,~B.; Wang,~J.; Kao,~A.; Pericleous,~K.; Phillion,~A.~B.; Atwood,~R.~C.;
  Lee,~P.~D. {4D synchrotron X-ray tomographic quantification of the transition
  from cellular to dendrite growth during directional solidification}.
  \emph{Acta Materialia} \textbf{2016}, \emph{117}, 160--169\relax
\mciteBstWouldAddEndPuncttrue
\mciteSetBstMidEndSepPunct{\mcitedefaultmidpunct}
{\mcitedefaultendpunct}{\mcitedefaultseppunct}\relax
\EndOfBibitem
\bibitem[Chang \latin{et~al.}(2012)Chang, Chen, Li, Yue, and Jin]{CHANG2012}
Chang,~G.-W.; Chen,~S.-Y.; Li,~Q.-C.; Yue,~X.-D.; Jin,~G.-C. {Lateral Growth
  Rate of Cells in Melt}. \emph{Journal of Iron and Steel Research,
  International} \textbf{2012}, \emph{19}, 28--50\relax
\mciteBstWouldAddEndPuncttrue
\mciteSetBstMidEndSepPunct{\mcitedefaultmidpunct}
{\mcitedefaultendpunct}{\mcitedefaultseppunct}\relax
\EndOfBibitem
\bibitem[Zhao \latin{et~al.}(2015)Zhao, Zhong, Huang, Ma, and Dong]{Zhao2015}
Zhao,~N.; Zhong,~Y.; Huang,~M.~L.; Ma,~H.~T.; Dong,~W. {Growth kinetics of
  Cu{\textless}inf{\textgreater}6{\textless}/inf{\textgreater}
  Sn{\textless}inf{\textgreater}5{\textless}/inf{\textgreater} intermetallic
  compound at liquid-solid interfaces in Cu/Sn/Cu interconnects under
  temperature gradient}. \emph{Scientific Reports} \textbf{2015}, \emph{5},
  1--12\relax
\mciteBstWouldAddEndPuncttrue
\mciteSetBstMidEndSepPunct{\mcitedefaultmidpunct}
{\mcitedefaultendpunct}{\mcitedefaultseppunct}\relax
\EndOfBibitem
\bibitem[Salloum-Abou-Jaoude \latin{et~al.}(2015)Salloum-Abou-Jaoude, Reinhart,
  Combeau, Zalo{\v{z}}nik, Lafford, and Nguyen-Thi]{Salloum-Abou-Jaoude2015}
Salloum-Abou-Jaoude,~G.; Reinhart,~G.; Combeau,~H.; Zalo{\v{z}}nik,~M.;
  Lafford,~T.~A.; Nguyen-Thi,~H. {Quantitative analysis by in situ synchrotron
  X-ray radiography of the evolution of the mushy zone in a fixed temperature
  gradient}. \emph{Journal of Crystal Growth} \textbf{2015}, \emph{411},
  88--95\relax
\mciteBstWouldAddEndPuncttrue
\mciteSetBstMidEndSepPunct{\mcitedefaultmidpunct}
{\mcitedefaultendpunct}{\mcitedefaultseppunct}\relax
\EndOfBibitem
\bibitem[Natarajan \latin{et~al.}(2014)Natarajan, Arivanandhan, Anandan,
  Sankaranarayanan, Ravi, Inatomi, and Hayakawa]{Natarajan2014}
Natarajan,~V.; Arivanandhan,~M.; Anandan,~P.; Sankaranarayanan,~K.; Ravi,~G.;
  Inatomi,~Y.; Hayakawa,~Y. {In-situ observation of faceted growth of
  benzophenone single crystals}. \emph{Materials Chemistry and Physics}
  \textbf{2014}, \emph{144}, 402--408\relax
\mciteBstWouldAddEndPuncttrue
\mciteSetBstMidEndSepPunct{\mcitedefaultmidpunct}
{\mcitedefaultendpunct}{\mcitedefaultseppunct}\relax
\EndOfBibitem
\bibitem[Fujiwara \latin{et~al.}(2011)Fujiwara, Gotoh, Yang, Koizumi, Nozawa,
  and Uda]{Fujiwara2011}
Fujiwara,~K.; Gotoh,~R.; Yang,~X.~B.; Koizumi,~H.; Nozawa,~J.; Uda,~S.
  {Morphological transformation of a crystal-melt interface during
  unidirectional growth of silicon}. \emph{Acta Materialia} \textbf{2011},
  \emph{59}, 4700--4708\relax
\mciteBstWouldAddEndPuncttrue
\mciteSetBstMidEndSepPunct{\mcitedefaultmidpunct}
{\mcitedefaultendpunct}{\mcitedefaultseppunct}\relax
\EndOfBibitem
\bibitem[Ye \latin{et~al.}(2017)Ye, Wang, Su, Chang, Wang, Zhang, Zhang, and
  Yang]{Ye2017}
Ye,~S.; Wang,~H.; Su,~H.; Chang,~L.; Wang,~S.; Zhang,~X.; Zhang,~J.; Yang,~B.
  {Facile fabrication of homogeneous and gradient plasmonic arrays with tunable
  optical properties via thermally regulated surface charge density}. \emph{J.
  Mater. Chem. C} \textbf{2017}, \emph{5}, 3962--3972\relax
\mciteBstWouldAddEndPuncttrue
\mciteSetBstMidEndSepPunct{\mcitedefaultmidpunct}
{\mcitedefaultendpunct}{\mcitedefaultseppunct}\relax
\EndOfBibitem
\bibitem[Jensen \latin{et~al.}(2013)Jensen, Lund, Gummel, Monkenbusch,
  Narayanan, and Pedersen]{Jensen2013}
Jensen,~G.~V.; Lund,~R.; Gummel,~J.; Monkenbusch,~M.; Narayanan,~T.;
  Pedersen,~J.~S. {Direct observation of the formation of surfactant micelles
  under nonisothermal conditions by synchrotron SAXS}. \emph{Journal of the
  American Chemical Society} \textbf{2013}, \emph{135}, 7214--7222\relax
\mciteBstWouldAddEndPuncttrue
\mciteSetBstMidEndSepPunct{\mcitedefaultmidpunct}
{\mcitedefaultendpunct}{\mcitedefaultseppunct}\relax
\EndOfBibitem
\bibitem[Jaeger and Chworos(2005)Jaeger, and Chworos]{Jaeger2005}
Jaeger,~L.; Chworos,~A. {From RNA Tectonics to Programmable Jigsaw Puzzles Made
  of RNA}. \emph{Foundations of Nanoscience. Edited by Reif JH. Science
  Technica} \textbf{2005}, 157--162\relax
\mciteBstWouldAddEndPuncttrue
\mciteSetBstMidEndSepPunct{\mcitedefaultmidpunct}
{\mcitedefaultendpunct}{\mcitedefaultseppunct}\relax
\EndOfBibitem
\bibitem[Talbot \latin{et~al.}(2017)Talbot, Kotar, Parolini, {Di Michele}, and
  Cicuta]{Talbot2017}
Talbot,~E.~L.; Kotar,~J.; Parolini,~L.; {Di Michele},~L.; Cicuta,~P.
  {Thermophoretic migration of vesicles depends on mean temperature and head
  group chemistry}. \emph{Nature Communications} \textbf{2017}, \emph{8},
  1--8\relax
\mciteBstWouldAddEndPuncttrue
\mciteSetBstMidEndSepPunct{\mcitedefaultmidpunct}
{\mcitedefaultendpunct}{\mcitedefaultseppunct}\relax
\EndOfBibitem
\bibitem[Selva \latin{et~al.}(2010)Selva, Miralles, Cantat, and
  Jullien]{Selva2010}
Selva,~B.; Miralles,~V.; Cantat,~I.; Jullien,~M.-C. {Thermocapillary actuation
  by optimized resistor pattern: bubbles and droplets displacing, switching and
  trapping}. \emph{Lab on a Chip} \textbf{2010}, \emph{10}, 1835\relax
\mciteBstWouldAddEndPuncttrue
\mciteSetBstMidEndSepPunct{\mcitedefaultmidpunct}
{\mcitedefaultendpunct}{\mcitedefaultseppunct}\relax
\EndOfBibitem
\bibitem[Balachandran \latin{et~al.}(2014)Balachandran, Panov, Panarin, Vij,
  Tamba, Mehl, and Song]{Balachandran2014}
Balachandran,~R.; Panov,~V.~P.; Panarin,~Y.~P.; Vij,~J.~K.; Tamba,~M.~G.;
  Mehl,~G.~H.; Song,~J.~K. {Flexoelectric behavior of bimesogenic liquid
  crystals in the nematic phase – observation of a new self-assembly pattern
  at the twist-bend nematic and the nematic interface}. \emph{J. Mater. Chem.
  C} \textbf{2014}, \emph{2}, 8179--8184\relax
\mciteBstWouldAddEndPuncttrue
\mciteSetBstMidEndSepPunct{\mcitedefaultmidpunct}
{\mcitedefaultendpunct}{\mcitedefaultseppunct}\relax
\EndOfBibitem
\bibitem[Basson and Pottebaum(2012)Basson, and Pottebaum]{Basson2012}
Basson,~M.; Pottebaum,~T.~S. {Measuring the temperature of fluid in a
  micro-channel using thermochromic liquid crystals}. \emph{Experiments in
  Fluids} \textbf{2012}, \emph{53}, 803--814\relax
\mciteBstWouldAddEndPuncttrue
\mciteSetBstMidEndSepPunct{\mcitedefaultmidpunct}
{\mcitedefaultendpunct}{\mcitedefaultseppunct}\relax
\EndOfBibitem
\bibitem[Schindelin \latin{et~al.}(2012)Schindelin, Arganda-Carreras, Frise,
  Kaynig, Longair, Pietzsch, Preibisch, Rueden, Saalfeld, Schmid, Tinevez,
  White, Hartenstein, Eliceiri, Tomancak, and Cardona]{Schindelin2012}
Schindelin,~J. \latin{et~al.}  {Fiji: an open-source platform for
  biological-image analysis.} \emph{Nature methods} \textbf{2012}, \emph{9},
  676--82\relax
\mciteBstWouldAddEndPuncttrue
\mciteSetBstMidEndSepPunct{\mcitedefaultmidpunct}
{\mcitedefaultendpunct}{\mcitedefaultseppunct}\relax
\EndOfBibitem
\bibitem[Arbeloa \latin{et~al.}(1989)Arbeloa, Ojeda, and Arbeloa]{Arbeloa1989}
Arbeloa,~F.; Ojeda,~P.; Arbeloa,~I. {Flourescence self-quenching of the
  molecular forms of Rhodamine B in aqueous and ethanolic solutions}.
  \emph{Journal of Luminescence} \textbf{1989}, \emph{44}, 105--112\relax
\mciteBstWouldAddEndPuncttrue
\mciteSetBstMidEndSepPunct{\mcitedefaultmidpunct}
{\mcitedefaultendpunct}{\mcitedefaultseppunct}\relax
\EndOfBibitem
\bibitem[Deville \latin{et~al.}(2013)Deville, Adrien, Maire, Scheel, {Di
  Michiel}, and Michiel]{Deville2013}
Deville,~S.; Adrien,~J.; Maire,~E.; Scheel,~M.; {Di Michiel},~M.;
  Michiel,~M.~D. {Time-lapse, three-dimensional in situ imaging of ice crystal
  growth in a colloidal silica suspension}. \emph{Acta Mater.} \textbf{2013},
  \emph{61}, 2077--2086\relax
\mciteBstWouldAddEndPuncttrue
\mciteSetBstMidEndSepPunct{\mcitedefaultmidpunct}
{\mcitedefaultendpunct}{\mcitedefaultseppunct}\relax
\EndOfBibitem
\bibitem[Marcellini \latin{et~al.}(2016)Marcellini, Noirjean, Dedovets, Maria,
  and Deville]{Marcellini2016}
Marcellini,~M.; Noirjean,~C.; Dedovets,~D.; Maria,~J.; Deville,~S. {Time-Lapse,
  in Situ Imaging of Ice Crystal Growth Using Confocal Microscopy}. \emph{ACS
  Omega} \textbf{2016}, \emph{1}, 1019--1026\relax
\mciteBstWouldAddEndPuncttrue
\mciteSetBstMidEndSepPunct{\mcitedefaultmidpunct}
{\mcitedefaultendpunct}{\mcitedefaultseppunct}\relax
\EndOfBibitem
\bibitem[Deville(2017)]{Deville2017}
Deville,~S. \emph{Freez. Colloids Obs. Princ. Control. Use}; 2017; pp
  1--46\relax
\mciteBstWouldAddEndPuncttrue
\mciteSetBstMidEndSepPunct{\mcitedefaultmidpunct}
{\mcitedefaultendpunct}{\mcitedefaultseppunct}\relax
\EndOfBibitem
\bibitem[Watanabe and Mizoguchi(2000)Watanabe, and Mizoguchi]{Watanabe2000}
Watanabe,~K.; Mizoguchi,~M. {Ice configuration near a growing ice lens in a
  freezing porous medium consisting of micro glass particles}. \emph{J. Cryst.
  Growth} \textbf{2000}, \emph{213}, 135--140\relax
\mciteBstWouldAddEndPuncttrue
\mciteSetBstMidEndSepPunct{\mcitedefaultmidpunct}
{\mcitedefaultendpunct}{\mcitedefaultseppunct}\relax
\EndOfBibitem
\bibitem[Dash \latin{et~al.}(1999)Dash, Fu, and Wettlaufer]{Dash1999}
Dash,~J.~G.; Fu,~H.; Wettlaufer,~J.~S. {The premelting of ice and its
  environmental consequences}. \emph{Reports Prog. Phys.} \textbf{1999},
  \emph{58}, 115--167\relax
\mciteBstWouldAddEndPuncttrue
\mciteSetBstMidEndSepPunct{\mcitedefaultmidpunct}
{\mcitedefaultendpunct}{\mcitedefaultseppunct}\relax
\EndOfBibitem
\bibitem[Hansen-Goos and Wettlaufer(2010)Hansen-Goos, and
  Wettlaufer]{HansenGoos2010}
Hansen-Goos,~H.; Wettlaufer,~J.~S. {Theory of ice premelting in porous media}.
  \emph{Phys. Rev. E} \textbf{2010}, \emph{81}, 31604\relax
\mciteBstWouldAddEndPuncttrue
\mciteSetBstMidEndSepPunct{\mcitedefaultmidpunct}
{\mcitedefaultendpunct}{\mcitedefaultseppunct}\relax
\EndOfBibitem
\bibitem[Coasne \latin{et~al.}(2013)Coasne, Galarneau, Pellenq, and {Di
  Renzo}]{Coasne2013}
Coasne,~B.; Galarneau,~A.; Pellenq,~R. J.~M.; {Di Renzo},~F. {Adsorption,
  intrusion and freezing in porous silica: the view from the nanoscale.}
  \emph{Chem. Soc. Rev.} \textbf{2013}, 4141--4171\relax
\mciteBstWouldAddEndPuncttrue
\mciteSetBstMidEndSepPunct{\mcitedefaultmidpunct}
{\mcitedefaultendpunct}{\mcitedefaultseppunct}\relax
\EndOfBibitem
\bibitem[Deville(2017)]{Deville2017g}
Deville,~S. \emph{Freez. Colloids Obs. Princ. Control. Use}; 2017; pp
  351--438\relax
\mciteBstWouldAddEndPuncttrue
\mciteSetBstMidEndSepPunct{\mcitedefaultmidpunct}
{\mcitedefaultendpunct}{\mcitedefaultseppunct}\relax
\EndOfBibitem
\bibitem[Deville \latin{et~al.}(2007)Deville, Saiz, and Tomsia]{Deville2007a}
Deville,~S.; Saiz,~E.; Tomsia,~A.~P. {Ice-templated porous alumina structures}.
  \emph{Acta Mater.} \textbf{2007}, \emph{55}, 1965--1974\relax
\mciteBstWouldAddEndPuncttrue
\mciteSetBstMidEndSepPunct{\mcitedefaultmidpunct}
{\mcitedefaultendpunct}{\mcitedefaultseppunct}\relax
\EndOfBibitem
\bibitem[Rempel and Worster(2001)Rempel, and Worster]{Rempel2001a}
Rempel,~A.~W.; Worster,~M.~G. {Particle trapping at an advancing solidification
  front with interfacial-curvature effects}. \emph{J. Cryst. Growth}
  \textbf{2001}, \emph{223}, 420--432\relax
\mciteBstWouldAddEndPuncttrue
\mciteSetBstMidEndSepPunct{\mcitedefaultmidpunct}
{\mcitedefaultendpunct}{\mcitedefaultseppunct}\relax
\EndOfBibitem
\bibitem[Liu \latin{et~al.}(2003)Liu, Muldrew, Wan, and Elliott]{Liu2003a}
Liu,~Z.; Muldrew,~K.; Wan,~R.~G.; Elliott,~J. A.~W. {Measurement of freezing
  point depression of water in glass capillaries and the associated ice front
  shape}. \emph{Phys. Rev. E} \textbf{2003}, \emph{67}, 61602\relax
\mciteBstWouldAddEndPuncttrue
\mciteSetBstMidEndSepPunct{\mcitedefaultmidpunct}
{\mcitedefaultendpunct}{\mcitedefaultseppunct}\relax
\EndOfBibitem
\bibitem[Lin \latin{et~al.}(2008)Lin, He, Dong, Liu, Xiao, and Liu]{Lin2008}
Lin,~C.; He,~G.; Dong,~C.; Liu,~H.; Xiao,~G.; Liu,~Y. {Effect of Oil Phase
  Transition on Freeze/Thaw-Induced Demulsification of Water-in-Oil Emulsions}.
  \emph{Langmuir} \textbf{2008}, \emph{24}, 5291--5298\relax
\mciteBstWouldAddEndPuncttrue
\mciteSetBstMidEndSepPunct{\mcitedefaultmidpunct}
{\mcitedefaultendpunct}{\mcitedefaultseppunct}\relax
\EndOfBibitem
\bibitem[Bogdan \latin{et~al.}(2016)Bogdan, Molina, and Tenhu]{Bogdan2016}
Bogdan,~A.; Molina,~M.~J.; Tenhu,~H. {Freezing and glass transitions upon
  cooling and warming and ice/freeze-concentration-solution morphology of
  emulsified aqueous citric acid}. \emph{Eur. J. Pharm. Biopharm.}
  \textbf{2016}, \emph{109}, 49--60\relax
\mciteBstWouldAddEndPuncttrue
\mciteSetBstMidEndSepPunct{\mcitedefaultmidpunct}
{\mcitedefaultendpunct}{\mcitedefaultseppunct}\relax
\EndOfBibitem
\bibitem[Ghosh and Coupland(2008)Ghosh, and Coupland]{Ghosh2008}
Ghosh,~S.; Coupland,~J.~N. {Factors affecting the freeze-thaw stability of
  emulsions}. \emph{Food Hydrocoll.} \textbf{2008}, \emph{22}, 105--111\relax
\mciteBstWouldAddEndPuncttrue
\mciteSetBstMidEndSepPunct{\mcitedefaultmidpunct}
{\mcitedefaultendpunct}{\mcitedefaultseppunct}\relax
\EndOfBibitem
\bibitem[Degner \latin{et~al.}(2014)Degner, Chung, Schlegel, Hutkins, and
  Mcclements]{Degner2014}
Degner,~B.~M.; Chung,~C.; Schlegel,~V.; Hutkins,~R.; Mcclements,~D.~J. {Factors
  influencing the freeze-thaw stability of emulsion-based foods}. \emph{Compr.
  Rev. Food Sci. Food Saf.} \textbf{2014}, \emph{13}, 98--113\relax
\mciteBstWouldAddEndPuncttrue
\mciteSetBstMidEndSepPunct{\mcitedefaultmidpunct}
{\mcitedefaultendpunct}{\mcitedefaultseppunct}\relax
\EndOfBibitem
\bibitem[Katsuki \latin{et~al.}(2017)Katsuki, Miyagawa, Nakagawa, and
  Adachi]{Katsuki2017}
Katsuki,~K.; Miyagawa,~Y.; Nakagawa,~K.; Adachi,~S. {Dispersion Stability of
  O/W Emulsions with Different Oil Contents Under Various Freezing and Thawing
  Conditions}. \emph{J. Food Sci.} \textbf{2017}, \emph{82}, 1569--1573\relax
\mciteBstWouldAddEndPuncttrue
\mciteSetBstMidEndSepPunct{\mcitedefaultmidpunct}
{\mcitedefaultendpunct}{\mcitedefaultseppunct}\relax
\EndOfBibitem
\bibitem[Chatterjee \latin{et~al.}(2016)Chatterjee, {Sen Gupta}, and
  Kumaraswamy]{Chatterjee2016}
Chatterjee,~S.; {Sen Gupta},~S.; Kumaraswamy,~G. {Omniphilic Polymeric Sponges
  by Ice Templating}. \emph{Chem. Mater.} \textbf{2016}, \emph{28},
  1823--1831\relax
\mciteBstWouldAddEndPuncttrue
\mciteSetBstMidEndSepPunct{\mcitedefaultmidpunct}
{\mcitedefaultendpunct}{\mcitedefaultseppunct}\relax
\EndOfBibitem
\bibitem[Khapli \latin{et~al.}(2009)Khapli, Kim, Montclare, Levicky, Porfiri,
  and Sofou]{Khapli2009}
Khapli,~S.; Kim,~J.~R.; Montclare,~J.~K.; Levicky,~R.; Porfiri,~M.; Sofou,~S.
  {Frozen cyclohexane-in-water emulsion as a sacrificial template for the
  synthesis of multilayered polyelectrolyte microcapsules.} \emph{Langmuir}
  \textbf{2009}, \emph{25}, 9728--9733\relax
\mciteBstWouldAddEndPuncttrue
\mciteSetBstMidEndSepPunct{\mcitedefaultmidpunct}
{\mcitedefaultendpunct}{\mcitedefaultseppunct}\relax
\EndOfBibitem
\bibitem[Dedovets \latin{et~al.}(2017)Dedovets, Monteux, and
  Deville]{Dedovets2017}
Dedovets,~D.; Monteux,~C.; Deville,~S. {Freezing Emulsions}. \emph{arXiv
  Prepr.} \textbf{2017}, \relax
\mciteBstWouldAddEndPunctfalse
\mciteSetBstMidEndSepPunct{\mcitedefaultmidpunct}
{}{\mcitedefaultseppunct}\relax
\EndOfBibitem
\bibitem[Deville(2017)]{Deville2017b}
Deville,~S. \emph{Freez. Colloids Obs. Princ. Control. Use}; 2017; pp
  1--46\relax
\mciteBstWouldAddEndPuncttrue
\mciteSetBstMidEndSepPunct{\mcitedefaultmidpunct}
{\mcitedefaultendpunct}{\mcitedefaultseppunct}\relax
\EndOfBibitem
\bibitem[Luo and Zhang(2012)Luo, and Zhang]{Luo2012}
Luo,~Z.; Zhang,~S. {Designer nanomaterials using chiral self-assembling peptide
  systems and their emerging benefit for society}. \emph{Chemical Society
  Reviews} \textbf{2012}, \emph{41}, 4736\relax
\mciteBstWouldAddEndPuncttrue
\mciteSetBstMidEndSepPunct{\mcitedefaultmidpunct}
{\mcitedefaultendpunct}{\mcitedefaultseppunct}\relax
\EndOfBibitem
\bibitem[Busseron \latin{et~al.}(2013)Busseron, Ruff, Moulin, and
  Giuseppone]{Busseron2013}
Busseron,~E.; Ruff,~Y.; Moulin,~E.; Giuseppone,~N. {Supramolecular
  self-assemblies as functional nanomaterials}. \emph{Nanoscale} \textbf{2013},
  \emph{5}, 7098\relax
\mciteBstWouldAddEndPuncttrue
\mciteSetBstMidEndSepPunct{\mcitedefaultmidpunct}
{\mcitedefaultendpunct}{\mcitedefaultseppunct}\relax
\EndOfBibitem
\bibitem[Hu \latin{et~al.}(2014)Hu, Gopinadhan, and Osuji]{Hu2014}
Hu,~H.; Gopinadhan,~M.; Osuji,~C.~O. {Directed self-assembly of block
  copolymers: a tutorial review of strategies for enabling nanotechnology with
  soft matter}. \emph{Soft Matter} \textbf{2014}, \emph{10}, 3867\relax
\mciteBstWouldAddEndPuncttrue
\mciteSetBstMidEndSepPunct{\mcitedefaultmidpunct}
{\mcitedefaultendpunct}{\mcitedefaultseppunct}\relax
\EndOfBibitem
\bibitem[Yang and Kotov(2011)Yang, and Kotov]{Yang2011}
Yang,~M.; Kotov,~N.~A. {Nanoscale helices from inorganic materials}.
  \emph{Journal of Materials Chemistry} \textbf{2011}, \emph{21}, 6775\relax
\mciteBstWouldAddEndPuncttrue
\mciteSetBstMidEndSepPunct{\mcitedefaultmidpunct}
{\mcitedefaultendpunct}{\mcitedefaultseppunct}\relax
\EndOfBibitem
\bibitem[Ariga \latin{et~al.}(2008)Ariga, Hill, Lee, Vinu, Charvet, and
  Acharya]{Ariga2008}
Ariga,~K.; Hill,~J.~P.; Lee,~M.~V.; Vinu,~A.; Charvet,~R.; Acharya,~S.
  {Challenges and breakthroughs in recent research on self-assembly}.
  \emph{Science and Technology of Advanced Materials} \textbf{2008},
  \emph{9}\relax
\mciteBstWouldAddEndPuncttrue
\mciteSetBstMidEndSepPunct{\mcitedefaultmidpunct}
{\mcitedefaultendpunct}{\mcitedefaultseppunct}\relax
\EndOfBibitem
\bibitem[Rogers \latin{et~al.}(2016)Rogers, Shih, and Manoharan]{Rogers2016}
Rogers,~W.~B.; Shih,~W.~M.; Manoharan,~V.~N. {Using DNA to program the
  self-assembly of colloidal nanoparticles and microparticles}. \emph{Nature
  Reviews Materials} \textbf{2016}, \emph{1}, 16008\relax
\mciteBstWouldAddEndPuncttrue
\mciteSetBstMidEndSepPunct{\mcitedefaultmidpunct}
{\mcitedefaultendpunct}{\mcitedefaultseppunct}\relax
\EndOfBibitem
\bibitem[Wang \latin{et~al.}(2016)Wang, Liu, Xing, and Yan]{Wang2016}
Wang,~J.; Liu,~K.; Xing,~R.; Yan,~X. {Peptide self-assembly: thermodynamics and
  kinetics}. \emph{Chemical Society Reviews} \textbf{2016}, \emph{45},
  5589--5604\relax
\mciteBstWouldAddEndPuncttrue
\mciteSetBstMidEndSepPunct{\mcitedefaultmidpunct}
{\mcitedefaultendpunct}{\mcitedefaultseppunct}\relax
\EndOfBibitem
\bibitem[Sorrenti \latin{et~al.}(2013)Sorrenti, Illa, and
  Ortu{\~{n}}o]{Sorrenti2013}
Sorrenti,~A.; Illa,~O.; Ortu{\~{n}}o,~R.~M. {Amphiphiles in aqueous solution:
  well beyond a soap bubble}. \emph{Chemical Society Reviews} \textbf{2013},
  \emph{42}, 8200\relax
\mciteBstWouldAddEndPuncttrue
\mciteSetBstMidEndSepPunct{\mcitedefaultmidpunct}
{\mcitedefaultendpunct}{\mcitedefaultseppunct}\relax
\EndOfBibitem
\bibitem[Liu \latin{et~al.}(2013)Liu, Zhou, Zhu, Yang, Liu, Wang, Zhang, and
  Zhuo]{Liu2013}
Liu,~Q.; Zhou,~H.; Zhu,~J.; Yang,~Y.; Liu,~X.; Wang,~D.; Zhang,~X.; Zhuo,~L.
  {Self-assembly into temperature dependent micro-/nano-aggregates of
  5,10,15,20-tetrakis(4-carboxyl phenyl)-porphyrin}. \emph{Materials Science
  and Engineering C} \textbf{2013}, \emph{33}, 4944--4951\relax
\mciteBstWouldAddEndPuncttrue
\mciteSetBstMidEndSepPunct{\mcitedefaultmidpunct}
{\mcitedefaultendpunct}{\mcitedefaultseppunct}\relax
\EndOfBibitem
\bibitem[Wang \latin{et~al.}(2015)Wang, Qi, Huang, Yang, Wang, Su, and
  He]{Wang2015}
Wang,~Y.; Qi,~W.; Huang,~R.; Yang,~X.; Wang,~M.; Su,~R.; He,~Z. {Rational
  Design of Chiral Nanostructures from Self-Assembly of a Ferrocene-Modified
  Dipeptide}. \emph{Journal of the American Chemical Society} \textbf{2015},
  \emph{137}, 7869−7880\relax
\mciteBstWouldAddEndPuncttrue
\mciteSetBstMidEndSepPunct{\mcitedefaultmidpunct}
{\mcitedefaultendpunct}{\mcitedefaultseppunct}\relax
\EndOfBibitem
\bibitem[Denkov \latin{et~al.}(2015)Denkov, Tcholakova, Lesov, Cholakova, and
  Smoukov]{Denkov2015}
Denkov,~N.; Tcholakova,~S.; Lesov,~I.; Cholakova,~D.; Smoukov,~S.~K.
  {Self-shaping of oil droplets via the formation of intermediate rotator
  phases upon cooling}. \emph{Nature} \textbf{2015}, \emph{528}, 392--395\relax
\mciteBstWouldAddEndPuncttrue
\mciteSetBstMidEndSepPunct{\mcitedefaultmidpunct}
{\mcitedefaultendpunct}{\mcitedefaultseppunct}\relax
\EndOfBibitem
\bibitem[Ziserman \latin{et~al.}(2011)Ziserman, Lee, Raghavan, Mor, and
  Danino]{Ziserman2011}
Ziserman,~L.; Lee,~H.~Y.; Raghavan,~S.~R.; Mor,~A.; Danino,~D. {Unraveling the
  mechanism of nanotube formation by chiral self-assembly of amphiphiles}.
  \emph{Journal of the American Chemical Society} \textbf{2011}, \emph{133},
  2511--2517\relax
\mciteBstWouldAddEndPuncttrue
\mciteSetBstMidEndSepPunct{\mcitedefaultmidpunct}
{\mcitedefaultendpunct}{\mcitedefaultseppunct}\relax
\EndOfBibitem
\bibitem[Aim{\'{e}} \latin{et~al.}(2007)Aim{\'{e}}, Manet, Satoh, Ihara, Park,
  Godde, and Oda]{Aime2007}
Aim{\'{e}},~C.; Manet,~S.; Satoh,~T.; Ihara,~H.; Park,~K.~Y.; Godde,~F.;
  Oda,~R. {Self-assembly of nucleoamphiphiles: Investigating nucleosides effect
  and the mechanism of micrometric helix formation}. \emph{Langmuir}
  \textbf{2007}, \emph{23}, 12875--12885\relax
\mciteBstWouldAddEndPuncttrue
\mciteSetBstMidEndSepPunct{\mcitedefaultmidpunct}
{\mcitedefaultendpunct}{\mcitedefaultseppunct}\relax
\EndOfBibitem
\bibitem[Aim{\'{e}} \latin{et~al.}(2009)Aim{\'{e}}, Tamoto, Satoh, Grelard,
  Dufourc, Buffeteau, Ihara, and Oda]{Aime2009}
Aim{\'{e}},~C.; Tamoto,~R.; Satoh,~T.; Grelard,~A.; Dufourc,~E.~J.;
  Buffeteau,~T.; Ihara,~H.; Oda,~R. {Nucleotide-promoted morphogenesis in
  amphiphile assemblies: Kinetic Control of micrometric helix formation}.
  \emph{Langmuir} \textbf{2009}, \emph{25}, 8489--8496\relax
\mciteBstWouldAddEndPuncttrue
\mciteSetBstMidEndSepPunct{\mcitedefaultmidpunct}
{\mcitedefaultendpunct}{\mcitedefaultseppunct}\relax
\EndOfBibitem
\end{mcitethebibliography}

\end{document}